\documentclass[floats,floatfix,showpacs,amssymb,prd,twocolumn,superscriptaddress,reprint,nofootinbib,nolongbibliography]{iopconfser}
\usepackage{orcidlink}

\usepackage{amssymb,amsmath, verbatim,mathtools,needspace,enumitem,etoolbox,graphicx, physics,microtype,afterpage,bigints,gensymb,tabularx,xspace, bm}
\usepackage{hyperref}
\definecolor{linkcolor}{rgb}{0.0,0.3,0.5}
\definecolor{urlcolor}{rgb}{0.27,0.55,0.}
\definecolor{funcolor}{rgb}{0.65, 0.16, 0.16}
\hypersetup{colorlinks=true,linkcolor=linkcolor,citecolor=linkcolor,filecolor=linkcolor,urlcolor=linkcolor}

\usepackage{titlesec}
\titlespacing*{\section}{0pt}{2ex}{1ex}

\usepackage[compress]{cite}



\begin{document}

\title{{\sc{precession 2.1}}: black-hole binary spin precession on eccentric orbits}

\author{
Giulia Fumagalli$\,$\orcidlink{0009-0004-2044-989X}$^{1,2}$, 
Davide Gerosa$\,$\orcidlink{0000-0002-0933-3579}$^{1,2}$, 
Nicholas Loutrel$\,$\orcidlink{0000-0002-1597-3281}$^{1,2}$}

\affil{$^1$~Dipartimento di Fisica ``G. Occhialini'', Universit\'a degli Studi di Milano-Bicocca, Piazza della Scienza 3, 20126 Milano, Italy}
\affil{$^2$~NFN, Sezione di Milano-Bicocca, Piazza della Scienza 3, 20126 Milano, Italy}

\email{\href{mailto:g.fumagalli47@campus.unimib.it}{g.fumagalli47@campus.unimib.it}}
\begin{abstract}
We present version 2.1 of the public code {\sc precession}, a Python module for studying the post-Newtonian dynamics of precessing black hole binaries. In this release, we extend the code to handle eccentric orbits. This extension leverages the existing numerical infrastructure wherever possible, introducing a semi-automatic method to adapt circular-orbit functions to the eccentric case via a Python decorator. Additional new features include orbit- and precession-averaged evolutionary equations for the eccentricity, as well as revised expressions to convert between post-Newtonian separation and gravitational-wave emission frequency.
 \end{abstract}

\section{Introduction}
The detection of gravitational waves (GWs) emitted during the final orbits of black hole (BH) binaries enables precise measurements of the properties of these systems~\cite{2025ApJ...995L..18A}. Some parameters, such as the BH masses and spin magnitudes, remain constant throughout the binary’s evolution. Others, such as spin orientations and orbital eccentricity, evolve significantly over time~\cite{1964PhRv..136.1224P,1994PhRvD..49.6274A}. As a result, their values at the time of GW detection generally differ from those at BH formation. While the former class of parameters provides some insight into the origin of BH binaries, it is often the latter that carries the most discriminating information about the formation channels of these systems~\cite{2018PhRvD..98h4036G,2021ApJ...921L..43Z}. %
Connecting measured quantities to their values at formation \cite{2021ApJ...921L..43Z,2022PhRvD.105b4076M}, or, conversely, to evolve the predictions of population synthesis codes~\cite{2024PhRvD.110f3012F,2022MNRAS.516.2252O,2024PhRvD.110d3023K,2020ApJS..247...48K}
 to the regime probed by GW detectors, is essential to gain insight into the formation channels of BH binaries. This requires tracking the evolution of the angular momenta vectors as binaries evolve on (potentially eccentric) orbits~\cite{2015PhRvD..92f4016G,2023PhRvD.108b4042G,2023PhRvD.108l4055F}.
A multi-timescale treatment of BH-binary dynamics forms the foundation of the formalism presented in Ref.~\cite{2015PhRvD..92f4016G,2023PhRvD.108b4042G} and implemented in the numerical code {\sc precession}~\cite{2016PhRvD..93l4066G}. %
While originally designed to handle spinning BH binaries on quasi-circular orbits, we have recently
 extended {\sc precession} to the more general case of eccentric binaries. The formalism underlying this extension has already been published in Ref.\cite{2023PhRvD.108l4055F}. In this short contribution to the proceedings of the 2025 GR+Amaldi conference, 
 we summarize the new features of the {\sc precession} code, which can now seamlessly perform post-Newtonian (PN) evolutions of spinning, eccentric BH binaries. Our modifications fall into two main categories: those implementing the transformations required to extend the circular formalism to eccentric binaries (Sec.~\ref{s1}), and those that have no circular counterpart and are unique to the eccentric case (Sec.~\ref{s2}). This extension to eccentric binaries constitutes version 2.1 of {\sc precession}. The code is publicly available at \href{https://github.com/dgerosa/precession}{github.com/dgerosa/precession}, where users can also find detailed documentation and example tutorials.

\begin{figure*}[h!]
\centering
\includegraphics[width=0.9\textwidth]{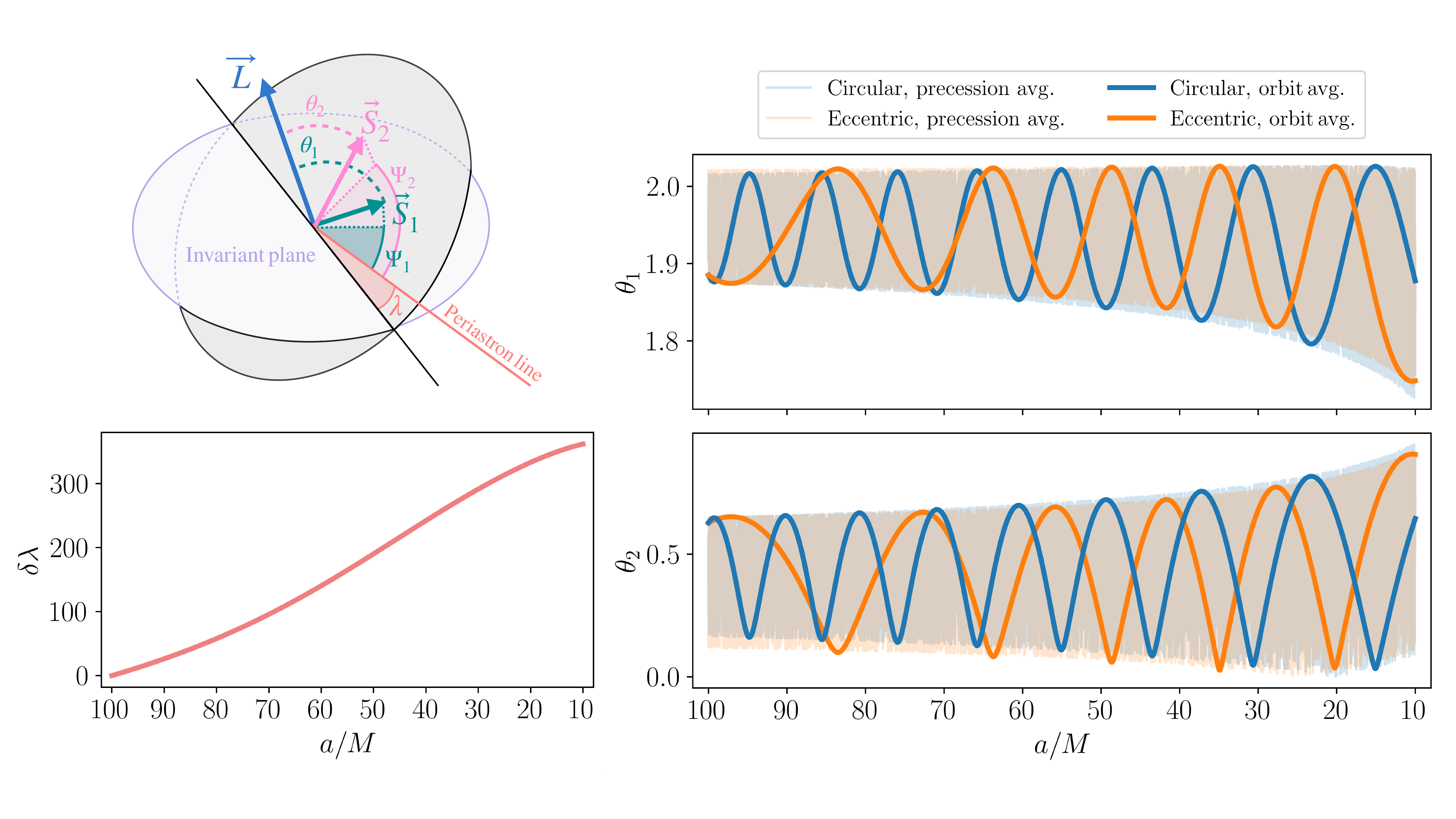} 
\caption{%
Orbit-averaged (heavy) and precession-averaged (light) evolution of the spin angles $\theta_1$ (top-right) and $\theta_2$ (bottom-right) under radiation reaction. The orbit-averaged periastron angle $\delta\lambda$ is shown in the bottom-left. We consider binaries with $q=0.8$, $\chi_1=0.7$, $\chi_2=0.6$, and initial $(\theta_1,\theta_2,\Delta\Phi)\approx (3/5\, \pi, 1/5\,\pi, 9/10\, \pi)$, evolved from $a_0=100M$ to $a=10M$ with $e_0=0$ (blue) and $e_0=0.5$ (orange). The top-left panel shows the geometric configuration.}
\label{fig}
\end{figure*}

\section{Mapping eccentric binaries to effective circular sources}\label{s1}
On the precessional timescale, the dynamics of an eccentric binary can be mapped onto those of an effective circular source~\cite{2023PhRvD.108l4055F}. The orbital separation (equal to the orbital radius $r$ in the circular case) is replaced by the semi-latus rectum, while the time coordinate $t$ must be rescaled by an eccentricity-dependent factor. Together, the mapping reads
\begin{align}
\label{e1}
a \longrightarrow a(1-e^2)\,, \qquad
t \longrightarrow t(1-e^2)^{3/2}\,,
\end{align}
where $a$ is the semi-major axis and $e$ is the orbital eccentricity, ranging from $0$ (circle) to $1$ (parabola). This transformation must be applied to all functions in the code that depend on $r$ and $t$, with a few exceptions (see Sec.~\ref{s2}). To implement the mapping %
 of the orbital separation, we developed a utility that dynamically wraps functions in {\sc precession} using a Python decorator. This decorator, named \texttt{eccentricize}, modifies the user-facing interface of selected functions by replacing the argument $r$ with $a$ and $e$. Internally, the functions continue to operate in terms of $r$, which is computed from $a$ and $e$ via Eq.~(\ref{e1}), thereby preserving their original logic while extending compatibility to eccentric binaries. The decorator also updates function signatures and docstrings to reflect the new input parameters. %
 Instead, functions that explicitly depend on time have been manually modified, as they %
  required additional adjustments %

\section{Functions requiring special treatment}\label{s2}
Not all functions in {\sc{precession}} can be extended to the eccentric case using the \texttt{eccentricize} decorator. Some require more careful treatment, either because they do not have a counterpart in the circular case (e.g., the definition and evolution of eccentricity) or because they involve non-trivial modifications.
The inspiral of a BH binary using orbit-averaged equations is implemented in \texttt{inspiral\_orbav}. For circular binaries, {\sc precession} employs the 2PN spin-precession equations together with the 3.5PN orbital-velocity evolution equation (including spin terms up to 2PN; see Sec.V of Ref.\cite{2023PhRvD.108b4042G}). For eccentric binaries, we have implemented the 2PN spin-precession equations and the 3PN evolutionary equations for the semi-major axis and eccentricity (with spin terms up to 2PN) [see Eqs. (C1a)–(C2l) of Ref.\cite{2018PhRvD..98j4043K}]. Three additional angular variables appear in this case: $\Psi_1$, $\Psi_2$, and $\delta\lambda$\cite{2019PhRvD.100l4008P,2018PhRvD..98j4043K,2021arXiv210610291K}. The first two describe the angles between the projections of the spin vectors onto the orbital plane and the periastron line, while the third identifies the position of the periastron, as illustrated in Fig.\ref{fig}. Therefore, in addition to tracking the spin orientations, it is also necessary to evolve $\delta\lambda$. In this version of the code, we adopt the evolution equation given in Eq.(17b) of Ref.\cite{2021arXiv210610291K}.
The precession-averaged evolution of eccentric binaries (\texttt{inspiral\_precav}) has been thoroughly described in Ref.~\cite{2023PhRvD.108l4055F}, to which we refer for further details. In particular, we have derived a new regularized version of the Peters' equation for $\dd a/\dd e$ that is well-suited to numerical integrations in the $e\to 0 $ limit. 
Figure~\ref{fig} shows an example of an eccentric BH binary evolution. We present the orbit-averaged evolution of the tilt angles $\theta_1$, $\theta_2$, and $\delta\lambda$ for a binary with initial eccentricity $e_0=0.5$, initialized at a large separation ($a=100M$) and evolved down to $a=10M$. For comparison, we also show the evolution of the spin angles for the same binary initialized with $e_0=0$ (the circular case). We note that, when setting $e=0$, the eccentric and circular evolutions do not coincide, because the derivative of $e$ %
 is not zero. This is a known feature of orbit-averaged equations~\cite{2010PhRvD..81l4001K,2018PhRvD..98j4043K,2019PhRvD.100l4008P}, which encodes the influence of spins on eccentricity. %
Another aspect that required careful handling is the conversion between the GW emission frequency $f_{\rm GW}$ and the orbital separation, namely \texttt{pnseparation\_to\_gwfrequency} and \texttt{gwfrequency\_to\_pnseparation}.
For eccentric binaries, GW emission is more complex than in the circular case~\cite{1963PhRv..131..435P}, as it occurs across multiple harmonics of the orbital frequency. Moreover, the distribution of power among these harmonics depends sensitively on the eccentricity~\cite{1963PhRv..131..435P, 2003ApJ...598..419W,2021RNAAS...5..275H}. In the version of the code presented here, we proceed as follows: given the semi-major axis $a$ and eccentricity $e$, we first compute the orbital frequency $\omega$ and then map $\omega$ to $f_{\rm GW}$, either considering the second harmonic, which dominates in the circular limit, or the harmonic that carries the maximum power. The latter is identified using the fitting formulae of Refs.~\cite{2003ApJ...598..419W,2021RNAAS...5..275H}, which we have reimplemented.
If $\omega$ or $f_{\rm GW}$ at a specific harmonic is known, and the eccentricity $e$ is given, the corresponding semi-major axis can be directly computed using Eqs.~(4a) and (B2a) of Ref.\cite{2018PhRvD..98j4043K}. To perform the inverse operation—recovering the frequency from the semi-major axis—we analytically invert Eq.~(B2a) of Ref.~\cite{2018PhRvD..98j4043K}, obtaining%
\begin{align}
 M \omega&=  \left(\frac{M}{a}\right)^{3/2}+\mathcal{ A}_{1 \rm PN} \left(\frac{M}{a}\right)^{5/2}+\mathcal{ A}_{1.5 \rm PN} \left(\frac{M}{a}\right)^{3}+\mathcal{ A}_{2 \rm PN}  \left(\frac{M}{a}\right)^{7/2}
\end{align}
where the coefficients are defined as
\begin{align}
\mathcal{ A}_{1\,\mathrm{PN}} &= \frac{-1}{2 (1+q)^2 (1 - e^2)} \left\{3 + q(5 + 3q) - e^2 \left[9 + q (17 + 9 q)\right]\right\},\\
\mathcal{ A}_{1.5\,\mathrm{PN}} &= \frac{-1}{2 (1+q)^2 (1 - e^2)^{3/2}} \left\{
 \left[2 + 3q + 3e^2 (2 + q)\right] \chi_1 \cos \theta_1 + q \left[3 + 2q + 3 e^2 (1 + 2 q)\right] \chi_2 \cos \theta_2
\right\},\\ %
\mathcal{ A}_{2\,\mathrm{PN}} &=
 \frac{1}{8 (1+q)^4 \left(1-e^2\right)^2 } \Big \{
75-60 \sqrt{1-e^2}+\left(323-216 \sqrt{1-e^2}\right) q+\left(499-312 \sqrt{1-e^2}\right) q^2+ \nonumber \\
& +\,\left(323-216 \sqrt{1-e^2}\right) q^3+ \left(75-60 \sqrt{1-e^2}\right) q^4+e^2 \left[36+60 \sqrt{1-e^2}+ q \,\Big (38\,+ \right. \nonumber \\
& \left. +\, 216 \sqrt{1-e^2}\Big)-\left(2-312 \sqrt{1-e^2}\right) q^2+ \left(38+216 \sqrt{1-e^2}\right) q^3+12 \left(3+5 \sqrt{1-e^2}\right) q^4\right] + \nonumber \\
&+ e^4 \left(147+563 q+835 q^2+563 q^3+147 q^4\right)+3 (1+q)^2 \left(1+e^2\right) \left[2\, q \,\chi_1 \chi_2 \,(2 \cos \theta_1 \cos \theta_2+ \right. \nonumber \\
& \left.
-\cos \Delta\Phi \sin \theta_1 \sin \theta_2)+\left(3 \cos \theta_1^2-1\right) \chi_1^2+\left(3 \cos \theta_2^2-1\right) q^2 \chi_2^2\,\right]\Big\}, %
\end{align}
where $M=m_1+m_2$ is the total mass of the system, $q=m_2/m_1$ is the mass ratio ($m_2\leq m_1$), $\chi_1$, $\chi_2$ are the spin magnitudes, and $\Delta\Phi$ is the difference between the projection of spin vectors onto the orbital plane.  In the previous, circular-only version of {\sc{precession}}, the conversions between $r$ and $f_{\rm GW}$ were based on Eqs.~(4.5) and (4.13) of Ref.~\cite{1995PhRvD..52..821K}. However, these did not include the spin-induced quadrupole term, a contribution that enters at 2PN order~\cite{1998PhRvD..57.5287P, 2008PhRvD..78d4021R}. 
In the current release of the code, we have corrected the expressions for the circular case, now including this missing contribution. While the impact of this term in the frequency-separation conversion equation is expected to be small, it is nonetheless important for consistency and completeness in spin-precession modeling.  {\sc precession} also includes a legacy implementation of analytic fits for the post-merger BH mass, spin, and recoil~\cite{2016PhRvD..93l4066G}. These fits were developed only for quasi-circular binaries and, therefore, have not been ported to the eccentric version of the code.

{\sc precession} is distributed via git  (\href{https://github.com/dgerosa/precession}{github.com/dgerosa/precession}); installation is as easy as {\tt pip install precession}. 
Currently, the default version of the code remains that for quasi-circular binaries, as described in Ref.~\cite{2023PhRvD.108b4042G}; this is accessible via {\tt import precession}. The functionalities for studying eccentric binaries described here and in Ref.~\cite{2023PhRvD.108l4055F} are implemented in a dedicated submodule, which we recommend importing as {\tt import precession.eccentricity as precession}. %

\section*{Acknowledgments}
We thank Matteo Boschini for discussions. 
G.F., D.G., and N.L. are supported by 
ERC Starting Grant No.~945155--GWmining, 
Cariplo Foundation Grant No.~2021-0555, 
MUR PRIN Grant No.~2022-Z9X4XS, 
Italian-French University (UIF/UFI) Grant No.~2025-C3-386,
MUR Grant ``Progetto Dipartimenti di Eccellenza 2023-2027'' (BiCoQ),
and the ICSC National Research Centre funded by NextGenerationEU. 
D.G. is supported by MSCA Fellowship No.~101064542--StochRewind, MSCA Fellowship No.~101149270--ProtoBH and MUR Young Researchers Grant No. SOE2024-0000125.
Computational work was performed at CINECA with allocations 
through INFN and Bicocca, and at NVIDIA with allocations through the Academic Grant program.

\bibliographystyle{iopart-num_edited}
\bibliography{precessionshort}
\end{document}